\documentclass[twocolumn,times,tighten]{aastex631}

\newcommand*{\toreferee}{} 
\newcommand*{\torefereetwo}{} 
\newcommand*{\KW}{} 

\usepackage{amsmath}
\usepackage{physics}
\usepackage{bbold}
\usepackage{soul}
\usepackage{booktabs} 
\usepackage{multirow} 
\usepackage{amsmath}  
\usepackage{lineno}

\graphicspath{{./}{figures/}}

\begin{document}



\title{Effects of Dust Coagulation on Streaming Instability}

\correspondingauthor{Ka Wai Ho}
\email{kho33@wisc.edu / kawaiho@lanl.gov}

\author[0000-0003-3328-6300]{Ka Wai Ho}
\affiliation{Theoretical Division, Los Alamos National Laboratory}
\affiliation{Department of Astronomy, University of Wisconsin-Madison}

\author{Hui Li}
\affiliation{Theoretical Division, Los Alamos National Laboratory}

\author{Shengtai Li}
\affiliation{Theoretical Division, Los Alamos National Laboratory}

\begin{abstract}%
Streaming Instability (SI) in dust has long been thought to be a promising process in triggering 
planetesimal formation in the protoplanetary disks (PPDs). In this study, we present the first numerical 
investigation that models the SI in the vertically stratified disk together with the dust coagulation process. 
Our simulations reveal that, even with the initial small dust sizes, because dust coagulation promotes 
dust size growth, SI can eventually still be triggered.
{\torefereetwo Specifically, dust coagulation, limited only by dust fragmentation, broadens the parameter boundaries obtained from previous SI studies using single dust species.}
We describe the various stages of dust dynamics along with their size evolution, and
explore the impact of different dust fragmentation velocities. 
Implications of these results for realistic PPDs are also discussed.   
\end{abstract}


\section{Introduction} \label{sec:intro}
The growth of micron-sized dust to kilometer-sized planetesimals is a key stage of the 
planet formation in PPDs. Various processes contribute to the overall evolution, 
including dust size growth and self-gravitational collapse of dust clumps. 
The latter process is expected to occur when
the concentrated dust clump density exceeds the critical 
Roche density. 
Several instability mechanisms  have been proposed to account for the accumulation stage of the dust clumps to trigger the 
self-gravitational collapse \citep{nas2022origins}, with the Streaming Instability (SI) being a prime candidate \citep{SI}.

{\toreferee
Extensive analytical and numerical studies have explored the conditions under which SI triggers the formation of sufficiently dense clumps to surpass the local Roche density
\citep[e.g.,][]{Johansen07,Kowalik13,Armitage16}. These studies have established
constraints on three critical parameters that control the clumping process:
the normalized stopping time ($\tau_s$), a measure of dust particle coupling to the gas;
the dust-to-gas (D/G) ratio; and the turbulent intensity
\citep{2020ApJ...895....4U,Chen20,LY21}. These parameters collectively determine
the conditions required for robust clumping.
Recent  investigations (largely axisymmetric) by \citet{LY21} have revealed that clumping can occur at even smaller particle sizes and lower D/G ratios than previously thought.}
Their findings show that clumping can arise with $D/$G ratios as low as 0.003 for
optimally-sized dust particles, or with $\tau_s$ as small as $10^{-3}$ for higher $D/G$ ratios. {\toreferee These results were obtained from extended simulations spanning 500 Keplerian orbits,
crucial for allowing the onset of strong SI clumping to fully develop across diverse parameters with effectively weak values of self-generated turbulence.}
{\torefereetwo Moreover, it remains uncertain whether the particle clumping behavior observed in 3D simulations is consistent with the results from axisymmetric studies
}

Most of these previous studies have used a single dust size (or a single $\tau_s$). 
{\KW In realistic PPDs, however, dust tends to have a size distribution along with
dust coagulation and fragmentation processes \citep[e.g.,][]
{2010AnA...513A..79B,2019ApJ...886...62L,2019ApJ...885...91D,2020ApJ...889L...8L,2020ApJ...892L..19L}. 
The dust size evolution opens up the possibility that the SI can eventually
get triggered in dusty environment where the initial dust sizes are relatively small
(or $\tau_s$ below the traditional SI threshold). In addition, 
there could be interplay between the evolution of SI and the dust sizes,
potentially altering SI dynamics and impacting dense clump formation,
as well as altering the dust size distribution and affecting the growth of larger particles.}
This is a topic of great interest \citep[e.g.,][]{D14, Schaffer18, McNally21, Tominaga24},
but due to the expensive computational cost of resolving the coagulation at high resolution
for extended long duration, most of the previous studies have focused on either analytical approach 
or simulations with a limited number of dust species that omit coagulation 
\citep{Schaffer18, Zhu21, Rucska23}. 
However, recent advancements in 
{\KW numerical simulations}, particularly the mesh refinement capabilities, 
significantly reduce the computational cost while maintaining the high resolution necessary to 
resolve the SI dynamics at the disk midplane.


In this paper, we present some of the first numerical simulations
incorporating both dust size evolution and the SI, aiming to explore one key question: 
{\KW how the dust size evolution will affect the SI. }
Our paper is structured as follows: Section \ref{sec:methods} describes 
our numerical scheme, Section \ref{sec:results} presents our results, 
Section \ref{sec:disc} discusses their implications, 
and Section \ref{sec:conclusions} summarizes our findings.

\section{Methods} \label{sec:methods}
\subsection{Hydrodynamical Modelling}

To simulate the coupled dynamics of gas and dust within a PPD, 
we employ the \texttt{Athena++} magnetohydrodynamic (MHD) code augmented 
with its integrated multi-fluid dust module. This configuration allows for the simultaneous 
evolution of both gas and dust species \citep{Athena++, AthenaDust}. 
{\KW An important distinction is that we have implemented a new multi-fluid approach 
for dust modeling along with dust coagulation dynamics in \texttt{Athena++}. 
To model dust coagulation and fragmentation, 
we adopt the implementation described in another code called LA-COMPASS
\citep{2019ApJ...886...62L,2019ApJ...885...91D,2020ApJ...889L...8L,2020ApJ...892L..19L}. 
which has a dust coagulation module using an explicit integration scheme to 
solve the Smoluchowski equation
 \cite{Brauer08}. This model incorporates turbulent mixing and Brownian 
motion as sources of collision velocities. To manage the high computational cost 
associated with solving the dust coagulation equations, we implement a 
sub-stepping routine. The dust coagulation module is called every $\Delta t = 0.1/ \Omega$ of the simulation 
, with $\Omega$ being the Keplerian frequency at radius $r$, ensuring a balance between computational efficiency and accuracy in the coagulation calculations. We employ a first-order implicit method for drag modeling in conjunction  with a second-order piecewise linear method to accurately  model the dynamic interactions between gas and dust. (The details of this new module in \texttt{Athena++} will be presented 
in a future publication.)
}


\begin{table}[hbt]
\centering
\caption{2D Simulation Parameters with Coagulation}
\label{tab:sim}
\begin{tabular}{cccccc}
\toprule
Runs & \(\tau_{s,0}\) & $\tau_s$ ranges & \(N_{dust}\)& \(v_f/c_s\)& \(\Omega t_{sim}\)  \\
\midrule
run1    & \(10^{-2}\) & $ [7.36\times10^{-6},0.736]$ &  101 & $2.5\times 10^{-3}$   & 2200 \\
run2    & \(10^{-2}\) & $ [7.36\times10^{-5},7.36]$ &  101 &  $5.0\times 10^{-3}$    &   2200 \\
run3   & \(10^{-2}\) &$ [2.58\times10^{-3},25.8]$ & 81 &  $2.5\times 10^{-2}$   &     2200 \\
run4  & \(10^{-3}\) & $ [2.58\times10^{-4},2.58]$ &  81 &  $2.5\times 10^{-2}$   &  3000 \\
\bottomrule
\end{tabular}
\end{table}


\subsection{Dust Evolution Modelling}

For dust, 
we assume it is in the Epstein's regime, so that Stokes Number 
$St = (\rho_p a v_K)/(\rho_g r c_s) \approx (a \rho_p)/\Sigma_g$, where
$\rho_p$ is the dust internal density (taken to be $1.25$ g/cm$^{3}$), $a$
is dust size, {\KW $c_s$ is the gas sound speed and $v_K$ is the gas Keplerian
speed at radius $r$. The disk scale height $H = r c_s/v_K$. 
The gas density is $\rho_g$ and its surface density is $\Sigma_g = \rho_g H$. }
Note that the variations in gas density during the simulations are very small.
The stopping time $t_s$ is normalized as 
$\tau_s =  t_s \Omega = St$. 
A key remaining parameter is the fragmentation velocity $v_f$. 
Due to its high variability and dependence on environmental factors like the 
presence of water ice, we investigate cases with 
{\KW $v_f/c_s = 2.5\times 10^{-3}, 5\times 10^{-3}$, and $2.5\times 10^{-2}$}
to explore how this parameter influences the SI.

To study the effects of dust coagulation on the SI, we follow the previous 
approach in numerical studies of the SI by utilizing a local shearing 
box approximation to model a vertically stratified patch of the PPD without self-gravity. 
Our physical setup and boundary conditions directly follow those defined in \cite{LY21}.
{\KW The setup employs a two-dimensional $x-z$ shearing box with 
the standard shearing periodic boundary conditions in $x$, 
and the outflow boundary condition in $z$.}
The simulation box has $L_x \times L_z = 0.8 H\times 0.4 H$, which has a 
base resolution of $240\times 480$. 
Furthermore, to reduce the computational cost, we introduce a specific mesh refinement 
at the midplane between $z= \pm 0.05 H$. This region has a resolution 
of $1920 \times 3840$ which is 
eight times higher than regions at higher disk heights. 
{\KW This will help capture most of the relevant dynamics occurring} at the midplane, 
and it ensures that the resolution requirements proposed by \cite{LY21} will be met.

Table \ref{tab:sim} lists the simulation parameters. 
{\KW For all these runs, the initial total dust to gas density ratio is set to be $D/G = 0.015$, which is to match the desired value from \cite{LY21}, for which the SI is not excited. }

For the initial dust size distribution, the dust at midplane initially has  
$\tau_{s,0} = 10^{-2}$ or $10^{-3}$ (see Table \ref{tab:sim}). 
{\KW To capture the dust size evolution across a wide range, we use either 81 or 101 logarithmic dust size bins, spanning four to five orders of magnitude in stopping time (see Table \ref{tab:sim}), ensuring a comprehensive representation of the dust population. }

{\KW To put these dimensionless parameters in perspective, we can use a 
minimum mass solar nebula model density profile located at 10 AU, 
$\Sigma_g \sim 53.3$ g/cm$^{2}$ and $c_s = 0.4$ km/s or $c_s/v_K = 0.04$. 
From these, we can find the gas density $\rho_g \approx
8.88\times 10^{-12}$ g/cm$^3$ and the Roche density is $\sim 5\times 10^{-10}$ g/cm$^3$,
which is $\sim 56\times$ the gas density. 
To study the interplay between vertical settling and dust coagulation effect, 
we set the initial dust height $H_d = 0.09 H$. {\toreferee The turbulence would be 
developed due to the excitation of SI, and no external sources of turbulence 
have been implemented.} The fragmentation velocity to 
sound speed ratio $v_f/c_s = 2.5, 5,$ and $25 \times 10^{-3}$ will give values of 
$v_f$ as $1, 2,$ and $10$ m/s, respectively. The initial dust size is $0.43$ cm for $\tau_{s,0} = 10^{-2}$. 
{\torefereetwo For all the simulations, We have used a parameter $\alpha_{coag}=10^{-5}$ in the coagulation model to mimic the turbulence expected from SI.} 
Furthermore, the dust sizes captured in simulations can vary between $\sim 3$ $\mu$m to $30$ cm for $\tau_s = 7.36\times 10^{-6}$ to $0.736$.}

For the computational cost, each simulation takes 5040 cores and ran for 2 to 4 days in NERSC Perlmutter or LANL Chicoma cluster, with an estimated cost of 200k to 300k CPU hours per run. 


\section{Results} \label{sec:results}
\subsection{Dust coagulation \& SI}

{\KW In the classical SI scenario, the development of SI can be divided 
into four stages: vertical settling, pre-clumping, strong clumping, and 
post-clumping phase. In the settling phase, dust particles vertically 
settle from higher latitudes to the mid-plane, resulting in an enhancement 
of mid-plane dust density. This is followed by the pre-clumping phase, 
a transition phase before strong clumping with a weak increase of mid-density density. 
In the strong clumping stage, it shows the excitation of SI with dust concentrations rapidly 
growing and merging into larger and denser clumps.  Finally, in the post-clumping 
stage, the largest dust aggregates that have reached Roche density decouple 
from the gas and have the potential to evolve into planetesimals through 
gravitational collapse. In this study, we will follow a similar classification 
scheme to better understand how dust coagulation makes an impact 
on the various stage of SI.
}



{\KW We emphasize that we have used an initial $D/G = 0.015$ for 
all the runs presented in this paper, which was shown not to 
excite SI in the single dust size study 
with $\tau_s \leq 10^{-2}$ by \cite{LY21}.}
We use this initial state to highlight how
dust coagulation could affect the overall evolution of 
dust clumping and whether SI can be excited after all. 
The basic idea is relatively straightforward: with coagulation, 
the dust size growth will increase their $\tau_s$, enhancing the 
coupling between dust and gas, and potentially leading to SI. 

\subsubsection{$\tau_{s,0} = 10^{-2}$ Cases}

\begin{figure}
    \centering
    \includegraphics[width=0.42\textwidth]{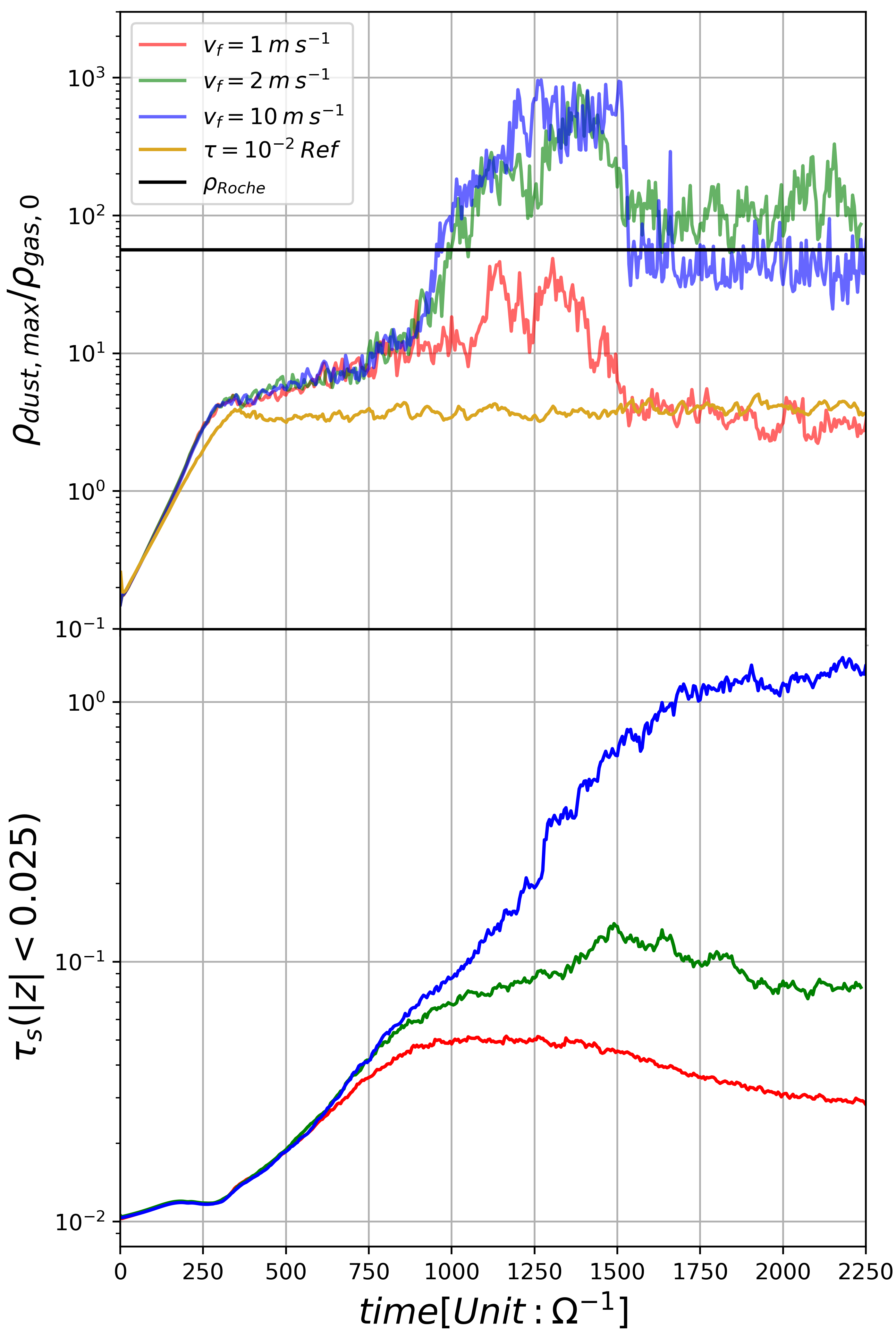}
    \caption{Temporal evolution of the mid-plane {\KW maximum} dust density (top) and the mass-averaged normalized stopping time (bottom) in simulations with the initial stopping time $\tau_{s,0} = 10^{-2}$ (the first three runs in Table \ref{tab:sim}). {\KW We adopt $c_s = 0.4$ km/s in all these runs.} The result from a single dust size run with $\tau_{s,0} = 10^{-2}$ without coagulation is also shown for reference (yellow lines).}
    \label{fig:1}
\end{figure}

Fig. \ref{fig:1} provides the evolution of the maximum dust clump density and 
the density-weighted averages of $\tau_s$ at the midplane region 
for all three runs with $\tau_{s,0} = 10^{-2}$ (see Table \ref{tab:sim}). 
It is evident that, in terms of the dust density evolution, 
all runs reach quite high maximum dust density
up to a thousand times the gas density, 
a clear sign of the SI. This outcome, which is different from the 
previous single fixed dust size simulation, is presumably solely 
caused by the dust size growth and is the key result of this paper. 

Similar to the previous single sized SI studies, these runs show approximately four stages: 
First, 
before $t < 300 \Omega^{-1}$, the dust 
vertically settles towards the mid-plane during the ``settling phase".
Second, the dust transitions into the ``pre-clumping phase” up to $t \sim 800 \Omega^{-1}$, 
with a relatively mild growth of the peak dust density by a factor of two.
Third, after $t > 800 \Omega^{-1}$, the dust enters the ``strong clumping phase", 
rather quickly for run2 and 3 though a bit slower for run1. 
The $\rho_{dust,max}$ grows quite robustly by another factory of 
up to $\sim 10^2$ for run2 and run3, and shows a weaker growth
by another large factor of $\sim 4$ for run1. 
This implies that the peak dust density will exceed the Roche density for run2
and approximately for run3,
but is still below it for run1. 
Finally, after these rigorous growth stages, all three runs transition into 
a ``quasi-equilibrium" stage between \(1250 \Omega^{-1}\) and \(1500 \Omega^{-1}\),
with a relatively steady maximum dust density. 
The final maximum dust density remains at several times the gas density for run1,
and $\sim 100$ times the gas density for run2 and 3. 


Along with the dust clumping and density growth, the dust size also grows 
in a corresponding fashion as shown in the bottom panel of Fig.\ref{fig:1}
where the density-weighted dust size within $z\pm 0.025$ is calculated. 
While the dust size growth is rather modest during the settling stage,
all three runs show similar amount of growth during the pre-clumping phase,
up to size $\tau_s \sim 0.4-0.6$. As the SI is excited and strong clumping ensures,
the growth during the strong clumping phase
is different for three runs, with higher fragmentation velocities reaching much
larger dust sizes. For example, run1 with the lowest $v_f$ shows a saturated 
dust size at $\tau_s \sim 0.05$, whereas run2 produces $\tau_s \sim 0.1$
and runs grows to $\tau_s \sim 0.8$. During the final equilibrium stage, 
the final $\tau_s$ for run1-3 settles at $0.03, 0.08,$ and $1.5$, respectively.           

It seems that the coagulation does not lead to any visible differences
among the three runs up to $t \sim 700 \Omega^{-1}$ but substantial differences
emerge after that. This is consistent with the expectation that, as higher density 
dust clumps develop due to SI, the coagulation process inside the dense clumps 
will alter the dust size distribution significantly. 

{The result of a single sized dust 
$\tau_{s,0} = 10^{-2}$ without coagulation 
is also shown in the top panel of Fig. \ref{fig:1}. As discussed earlier,
this case does not produce SI \citep{LY21}, so the final dust density 
(though showing increase due to the vertical settling) remains relatively low. 
This further supports the main conclusion that the coagulation is
essential in triggering the SI. 
}

{\KW

\begin{figure}
    \centering
    \includegraphics[width=0.42\textwidth]{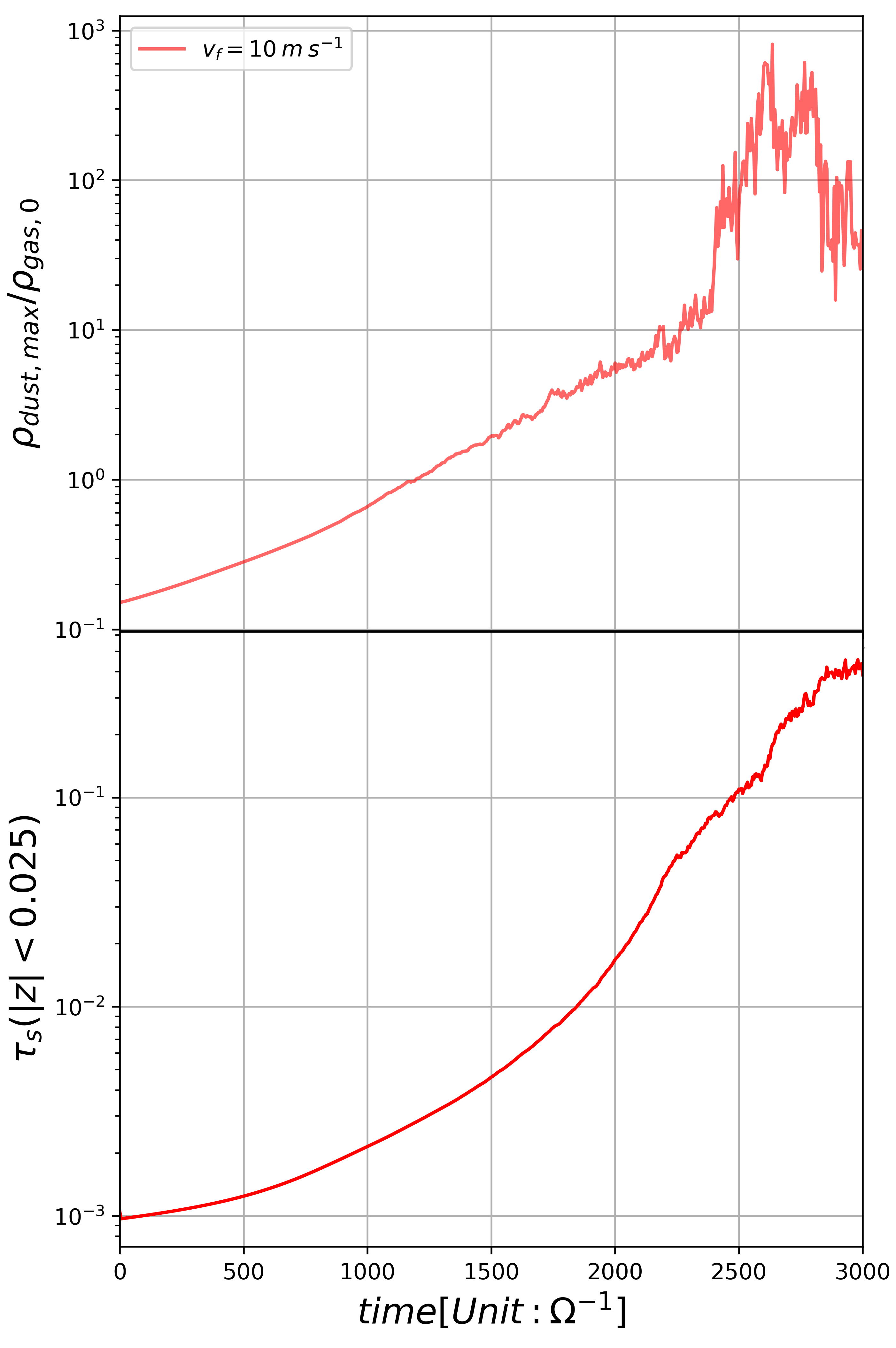}
    \caption{Temporal evolution for $\tau_{s,0}=10^{-3}$ case.}
    \label{fig:2}
\end{figure}

\subsubsection{$\tau_{s,0} = 10^{-3}$ Case}

Fig. \ref{fig:2} illustrates the evolution for $\tau_{s,0} = 10^{-3}$ and 
$v_f/c_s = 2.5\times 10^{-2}$. 
In contrast to the $\tau_{s,0} = 10^{-2}$ cases, run 4 demonstrates distinctive 
behavior during the vertical settling and pre-clumping phases. Because of the small initial
$\tau_{s}$, the vertical settling phase takes a lot longer, extending to 
$t \approx 2000 \Omega^{-1}$. Unlike the limited dust growth observed 
in $\tau_{s,0} = 10^{-2}$ cases, the dust size growth in this case is by a factor of 20 
during the settling phase. We also notice a slight dust density fluctuation 
starting at $t = 1500 \Omega^{-1}$, potentially indicating that some kind of 
instability in the dust may have already begun at this stage of the vertical settling phase. 
The system then transitions to the pre-clumping phase between $t = 2000$ 
and $2300 \Omega^{-1}$. The duration of the pre-clumping phase 
is also shorter compared to run 3. It continues
until the average dust size reaches about 0.08, 
at which point the system undergoes an instability, leading to a strong clumping 
phase with the peak density reaching $\sim 900 \rho_{g,0}$. 
Following this transition, the system's evolutionary trajectory closely 
aligns with that observed in run 3 with SI being excited.
Given its high $v_f$, the final $\tau_s$ is relatively large, reaching $\sim 0.4$.


}

\subsection{Mid-Plane Morphology of Dust Distribution}

\begin{figure*}[tbp]
    \centering
    \includegraphics[width=1.0\textwidth]{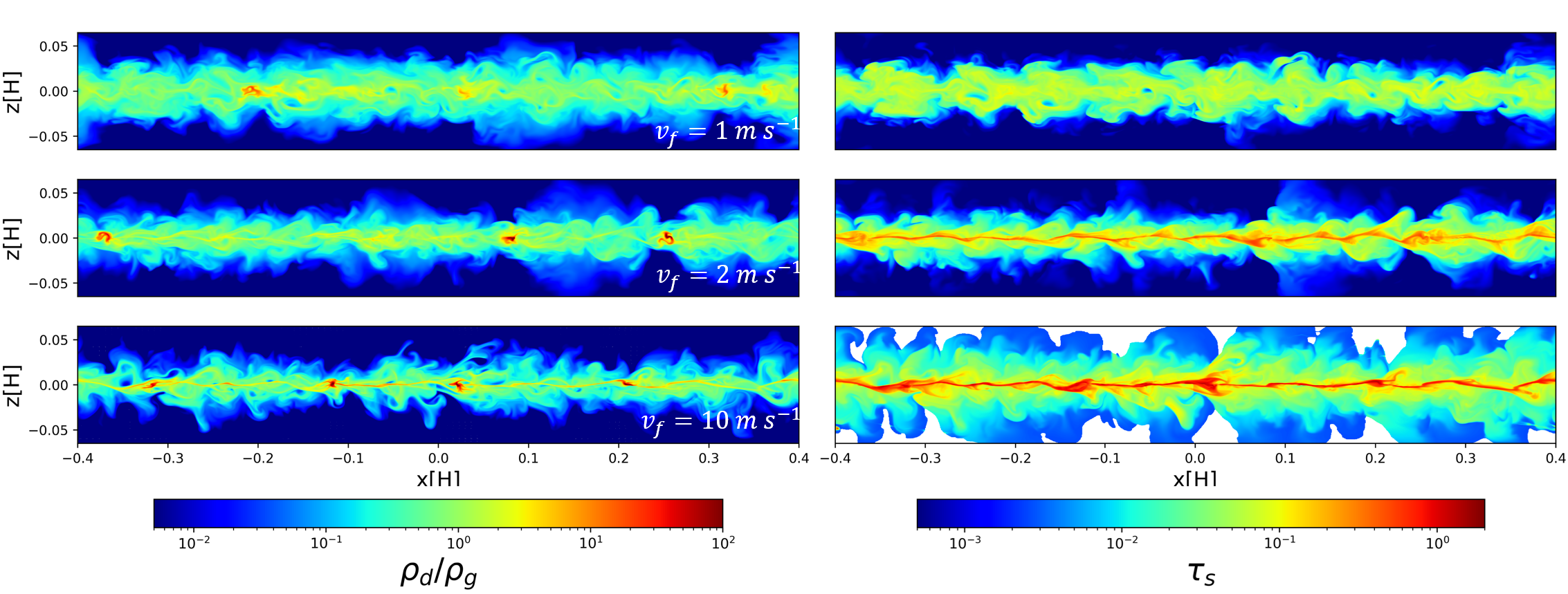}
    \caption{Spatial distribution of dust density ratio $\rho_d/\rho_g$ (left) and the mass-averaged stopping time 
    $\tau_s$ (right) for varying fragmentation velocities $v_f  = 1, 2,$ and $10$ m/s. (White region indicates local dust density falling below $10^{-8} \rho_g$ and the mass averaged statistics not computed.)
    Each frame is chosen to be $\sim 100 \Omega^{-1}$ after the 
strong clumping phase has started for run1-3 respectively (cf. Fig. \ref{fig:1}), which are $t = 1300 \Omega^{-1}$ (run 1), and $1350 \Omega^{-1}$(run 2 and 3).}
    \label{fig:3}
\end{figure*}

To understand the consequence of  coagulation on SI in more detail, 
Figure \ref{fig:3} displays snapshots of the dust density and 
mass-averaged stopping time, roughly $100 \Omega^{-1}$ after the 
strong clumping phase has started for run1-3 individually. 

All cases reveal the formation of dense dust clumps with 
turbulent structures from SI (the left panel of Fig. \ref{fig:3}). 
As the fragmentation velocity gets larger, 
several trends are visible. With higher $v_f$, dust is expected grow to larger sizes,
which is seen (the right panel of Fig. \ref{fig:3}). In turn, 
higher $\tau_s$ dust leads to a more spatially settled dust distribution 
towards the midplane. 
As more dust concentrates at the midplane, numerous highly concentrated dust clumps form. The highest density clumps also promote higher $\tau_s$ as well.





\subsection{Evolution of Dust Size Distribution}
\begin{figure*}[tbp]
    \centering
    \includegraphics[width=1.0\textwidth]{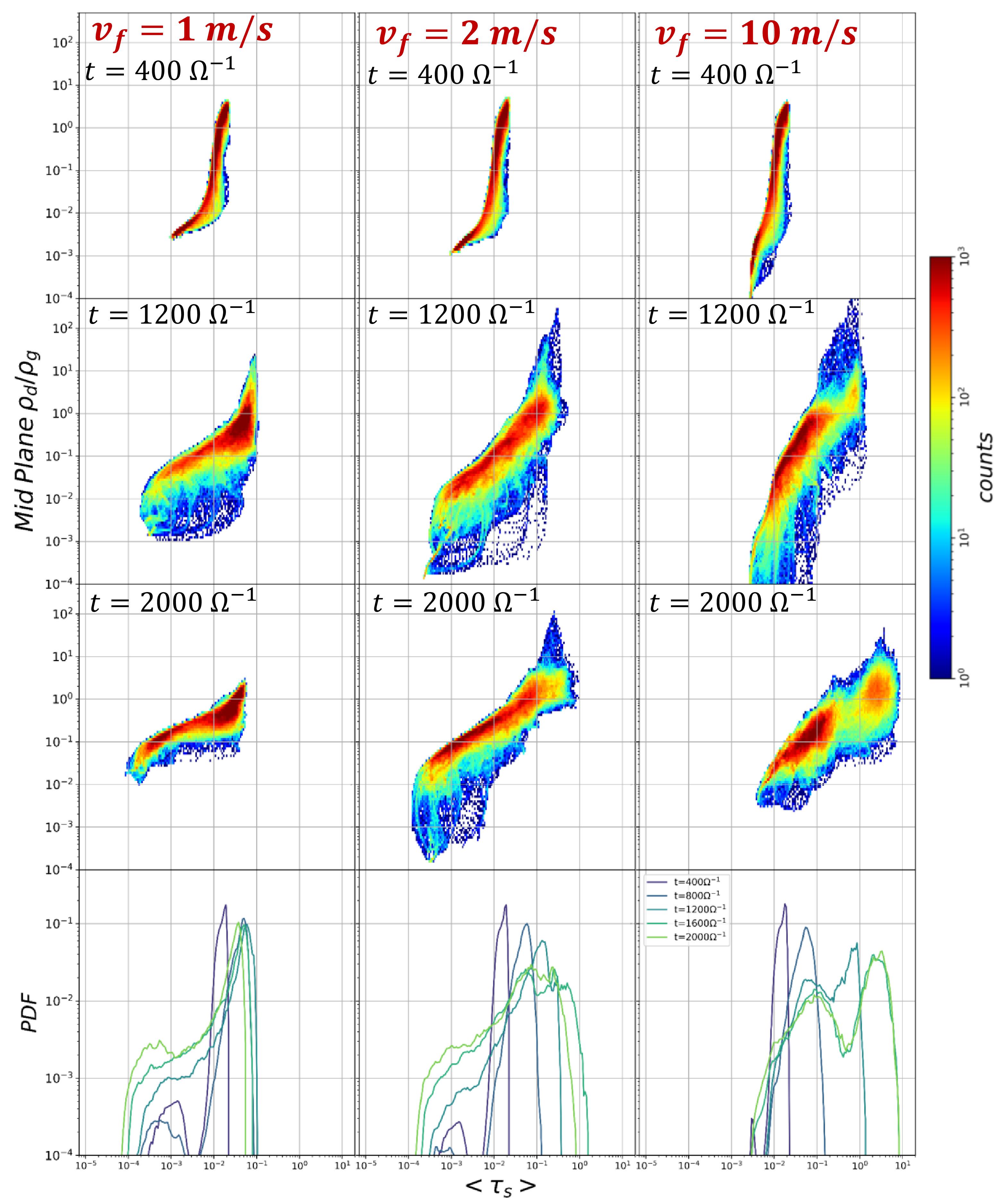}
    \caption{Temporal Evolution of the mid-plane ($|z| < 0.05 H$) dust density $\rho_d/\rho_g$ and its mass-weighted size 
    $\langle \tau_s \rangle$ for run 1-3 with three fragmentation velocities of $1 m/s$ (left column), 
    $2 m/s$ (middle), and $10 m/s$ (right), respectively. The bottom row is the evolution of the probability density function of dust size distribution for run 1-3, respectively. All runs have an initial stopping time 
    $\tau_{s,0}=10^{-2}$ and initial $\rho_d/\rho_g = 1.5\times 10^{-2}$. }
    \label{fig:4}
\end{figure*}

Figure \ref{fig:4} presents the joint evolution of dust density and size
(the top three rows) and the overall mass-weighted dust size
evolution (the bottom row), for run 1-3 with different fragmentation velocities. 
We emphasize that, for each of these runs, adequate dust size bins are used to allow
the dust size growth, as given in Table \ref{tab:sim}.
For all cases, by the end of the settling phase $\Omega t \sim 300-400$, 
there is only a modest dust size growth by a factor of $\sim 2$.
By the end of pre-clumping phase at $\Omega t \sim 800$ (bottom row),
the dust size distribution has now undergone significant 
coagulation and shifts its peak from the initial $\tau_s = 0.01$ to $\tau_s \sim 0.06$.
The similarity of dust size distributions for these three cases at this time
is consistent with the conclusions from Fig. \ref{fig:1} that the settling is faster
than coagulation for these runs and the coagulation mostly occur at the mid-plane
when the dust has settled. 

At $\Omega t = 1200$ and $2000$, which correspond to the
peak and saturation of SI respectively, different $v_f$ values make a marked difference.
For $v_f = 1$ m/s, the dust size distribution maintains its peak at $\tau_s \sim 0.06$ and 
the SI is visible but weak, producing relatively fewer dense dust clumps (cf. Fig. \ref{fig:3}).
For $v_f = 2$ m/s, the peak gradually increases beyond $\tau_s \sim 0.1$, and
shows signatures for a two-peak distribution at $\tau_s \sim 0.08 - 0.1$ and
$\tau_s \sim 0.3 - 0.5$, respectively. When combining with the spatial distributions 
shown in Fig. \ref{fig:1}, it is suggestive that the average dust size inside the densest 
clumps corresponds to the higher $\tau_s$ peak. 
For $10$ m/s, the dust size distributions eventually 
exhibit two peaks: one at $\tau_s \sim 0.1$
and the other at $\tau_s \sim 2-4$. We can see that the dust with $\tau_s \sim 2-4$ is
concentrated in a very narrow layer at the mid-plane along with the highest density clumps,
whereas dust with $\tau_s \sim 0.1$ occupies around the midplane with a 
wider extent.  
Evidently, for dust inside the dense clumps (due to SI), coagulation further
pushes their sizes to $\tau_s = 2-4$; for those dust that stay outside clumps, their
size (due to coagulation and fragmentation) is balanced at $\tau_s \approx 0.1$.
This eventually leads to the two dust size populations.
It is clear that larger $v_f$ promotes more dust grows into the larger sizes
and form a large number of clumps as well. 


\section{Discussion} \label{sec:disc}
{\KW
\textit{Relaxation of the Streaming Instability Threshold and 
the impact of coagulation in SI development.} Previous studies have found that the initiation of SI is contingent upon a specific range for both the average particle size and the mass ratio, which facilitates the mass accumulation of dust clumps. For example, for disks with a low mass ratio ($D/G < 0.02$), effective SI occurs within a particle size range of $\tau = 2\times 10^{-2}$ to $\tau = 1$ \citep{LY21}. 
A higher mass ratio enables smaller dust sizes to trigger SI. 
Our study demonstrates that the process of dust coagulation significantly  influences the SI dynamics. Notably, even for initial conditions with $\langle \tau_0 \rangle = 10^{-3}$, coagulation could increase the dust size by an order of magnitude within a few hundred orbits, thereby triggering the SI. This finding relaxes the more stringent particle size requirements previously identified (see Fig. 1 in \cite{LY21}). {\torefereetwo Due to the effect of dust coagulation, regions previously identified as SI-inactive could evolve into SI-active regimes within the timescale of SI evolution.}

Additionally, we observed differences in the evolution of SI between 
small and large initial dust size distributions. This is illustrated by 
comparing run 3 ($\tau_{s,0} = 10^{-2}$) and run 4 ($\tau_{s,0} = 10^{-3}$). 
The settling timescale can be estimated as 
$\tau_{set} \propto 1/\tau_{s,0}$ \citep{2014MNRAS.437.3055L} and 
the coagulation timescale is $\tau_{coag} \propto \tau_{s,0}$ \citep{Tominaga24}, assuming perfect sticking. On the one hand, for small $\tau_s$,  the coagulation timescale can be much shorter than the settling timescale, 
so the dust undergoes appreciable size growth during the settling process. On the other hand, it will take longer time for the initial small dust to grow
to a certain value of $\tau_s$ where SI can be triggered. 
The evolution trajectory shown in Fig. \ref{fig:2} indicates such dynamics. 

%
}

\textit{Importance of the Fragmentation Velocity.} Our results indicate that the 
fragmentation velocity, \( v_f \), is a critical factor in the dust growth process 
during the clumping phase. It influences both the peak dust-to-gas ratio and 
the resultant average dust size. In scenarios where \( v_f \) is low (e.g., \( v_f = 1 \) m/s), 
the coagulation process tends to be less efficient, which may kill the 
streaming instability during the strong clumping phase. Conversely, 
our findings show that a higher fragmentation velocity can promote both
dust size growth and effective dust clump formation. 


\textit{Conditions in Protoplanetary Disks.} 
The synergistic efficacy of SI and dust coagulation, as we have discussed, 
hinges on an extended time frame and a higher fragmentation velocity. 
A significant concern, however, is the radial drift of larger dust particles. 
Over the timescale of hundreds of orbits, this drift can lead to the 
inward drift of large dust grains, which is not completely captured in shearing box
simulations.
Moreover, the higher fragmentation velocity suggests a preferential composition 
of dust grains -- icy grains rather than silicates are advantageous. 
These conditions are likely to be found near the snow line, where the back-reaction 
effect can play a significant role \citep{2020A&A...635A.149G}. {\toreferee The snow line 
could act as a pressure bump,
impeding the inward drift of dust and potentially  creating a higher local dust-to-gas ratio region. {\torefereetwo Although the pressure bump effect of the snow line may not favor single-species SI for planetesimal formation in millimeter grin scenario  \citep{2022BAAS...54e4502C}, the presence of icy grains may elevate the fragmentation velocity threshold to 10 m/s, compared to the typical 1 m/s for silicate grains. This increase in fragmentation velocity promotes dust coagulation and growth, thereby sustaining SI during the strong clumping phase.
}}
Our studies has not included the effects of self-gravity, which is expected to play
a role particularly in the dense dust clumps. Such effects will be explored in future studies.

{\toreferee



\textit{Vertical Diffusion and Turbulence} 
{\toreferee There is the self-generated turbulence from SI, helping the vertical diffusion of gas and dust. The vertical diffusion can be estimated using $D_{p,z} = H_p^2 \Omega \tau_s$, or $\alpha_{diff} = D_{p}/{(c_s H)}$ where $H$ is the gas scale height,  $H_p$ is the mass-averaged dust scale height with $H \gg H_p$, and $\tau_s$ is the mass-averaged dust dimensionless stopping time, respectively \citep{1995Icar..114..237D, 2010EAS....41..187Y}. 
We evaluated the spatial and temporal evolution of 
$\alpha_{diff}$ for runs with initial $\tau_{s,0}=10^{-2}$. 
We find that the average $\alpha_{diff}$ at the mid-plane reaches
$\sim 10^{-5}$ during the nonlinear saturated stage of SI for both Run 1 and 2. 
For the highest $v_f = 10 m/s$ case (Run 3), the final $\alpha_{diff} \sim 10^{-4}$
as  $\tau_s$ approaches 1. These estimates are also consistent with our
analysis of Reynolds stresses using the actual gas velocity fluctuations in
three components \citep{1998RvMP...70....1B,2022BAAS...54e.381B}.
Generally the $\alpha_{diff}$ obtained in our study is 
consistent with the findings of \cite{2020ApJ...895....4U,LY21}. 
However, as noted by \cite{2023ApJ...942...74S}, the turbulence statistics of axis-symmetric simulations 
used in this study may differ from those in 3D simulations. Future studies using 3D simulations are needed.
%
%

In order to make direct comparisons with the previous non-coagulation studies of SI which did not include an explicit viscous processes in the hydrodynamics of gas and dust,  we opt not to include explicit viscosity in the hydrodynamics as well. {\torefereetwo As for the effects of turbulence on coagulation, a larger $\alpha_{coag}$ is expected to modify dust evolution significantly, as it would enhance the collision rate in the Smoluchowski equation. Consequently, stronger turbulence would lead to more rapid dust growth. However, the increased relative velocities associated with stronger turbulence would also impose a fragmentation barrier, limiting the maximum dust size that can be achieved. Future studies incorporating realistic $\alpha_{coag}$ values in multi-dimensional simulations are needed to better capture dust evolution under the influence of hydrodynamical motion.}
}
}

{\toreferee
\textit{The Bouncing Barrier and Its Impact on SI} 
The results presented here consider a dust coagulation model \citep{Brauer08}, accounting for the coagulation 
and fragmentation of dust grains. However, another important process in dust evolution, namely the bouncing physics,
has not been included \citep[e.g.,][]{2021NatRP...3..405W}. 
The bouncing barrier, where particles become less likely to stick together and instead tend to bounce off each other, could significantly alter the dust size distribution and, consequently, the dynamics of SI.
Previous studies on dust evolution models incorporating bouncing physics have primarily 
been conducted in one-dimensional simulations
\citep[e.g.,][]{2016ApJ...818..200E,2022ApJ...936...42E,2024Icar..41716085Y,2024A&A...682A.144D}. These studies suggest that the bouncing barrier can lead to a very narrow, size distribution, removing both large and $\mu$m-sized grains in the process, {\torefereetwo  with most dust particles growing and stabilizing at approximately $10^{-2}$ cm \citep{2024A&A...682A.144D}. }
It is currently unknown on how the inclusion of bouncing physics in our model  will alter the outcomes presented in this paper. If it leads to a more concentrated size distribution, it is conceivable that the effects of dust size growth on the SI development will be reduced as well.
%
However, it is important to note that dust evolution behavior can differ significantly in higher-dimensional simulations. 
This has been demonstrated for traditional dust coagulation models, where larger dust formation is more favored in 2D/3D settings \citep{2019ApJ...886...62L,2019ApJ...885...91D,2020ApJ...889L...8L,2020ApJ...892L..19L} compare to the 1D \citep{2012A&A...539A.148B}. Given these dimensional dependencies, the impact of the bouncing barrier on SI in multi-dimensional scenarios remains an open question, and the future multi dimensional simulations incorporating bouncing physics are needed.

While our current study does not include bouncing effects, it offers insights into certain aspects of particle growth and streaming instability under idealized conditions, providing a baseline for comparison with more complex models in future.
}

\section{Conclusions}  \label{sec:conclusions}
In this study, we have numerically explored the synergy between 
streaming instability and dust coagulation, specifically examining the 
strong particle clumping effect in the absence of self-gravity. 
Our findings emphasize the potential of dust size growth through coagulation 
to facilitate further the excitation and growth of SI. 
Our principal conclusions are:

\begin{enumerate}
    \item Within the timescale of a few hundred orbits, the dust coagulation could potentially grow the average dust size by a factor of several up to two orders of magnitudes, producing dust into $\tau_s \sim 0.1 - 1$ range, satisfying the condition for triggering SI. This extends the traditional boundary for exciting SI to even broader disk conditions. 


    \item The fragmentation velocity plays a role in regulating the dust size distribution in dense dust clumps and their surrounding regions. Consequently, it also affects the strength and final outcome of SI. 
    
%
\end{enumerate}

Numerical studies of the combined effects of the initial dust size, dust settling, coagulation, and SI 
are beginning to provide a more comprehensive picture of the dynamics in PPDs. Future studies
including the effects of {\torefereetwo dust bouncing}, snow lines and self-gravity will be crucial for understanding the planetesimal
formation.

\section*{Acknowledgments}
We acknowledge Mordecai-Mark Mac Low and Kratter Kaitlin for fruitful discussions during the final stage of the paper. We would also like to thank the referee for many useful comments and suggestions which significantly improved this work. We thank Michael Halfmoon for granting additional NERSC time needed for this project. Support by the LANL/LDRD program is gratefully acknowledged. This research used resources provided by the LANL Institutional Computing Program. This research also used resources of NERSC with award numbers FES-ERCAP-m4239 and m4364. 

\vspace{5mm}

\software{Athena++ \citep{Athena++}, MatPlotLib \citep{2007Matplotlib}, Julia \citep{Julia}, MHDFlows \citep{MHDFlows}}

\bibliographystyle{aasjournal}
\bibliography{refs}

\begin{thebibliography}{}
\expandafter\ifx\csname natexlab\endcsname\relax\def\natexlab#1{#1}\fi
\providecommand{\url}[1]{\href{#1}{#1}}
\providecommand{\dodoi}[1]{doi:~\href{http://doi.org/#1}{\nolinkurl{#1}}}
\providecommand{\doeprint}[1]{\href{http://ascl.net/#1}{\nolinkurl{http://ascl.net/#1}}}
\providecommand{\doarXiv}[1]{\href{https://arxiv.org/abs/#1}{\nolinkurl{https://arxiv.org/abs/#1}}}

\bibitem[{{Armitage} {et~al.}(2016){Armitage}, {Eisner}, \& {Simon}}]{Armitage16}
{Armitage}, P.~J., {Eisner}, J.~A., \& {Simon}, J.~B. 2016, \apjl, 828, L2, \dodoi{10.3847/2041-8205/828/1/L2}

\bibitem[{{Balbus} \& {Hawley}(1998)}]{1998RvMP...70....1B}
{Balbus}, S.~A., \& {Hawley}, J.~F. 1998, Reviews of Modern Physics, 70, 1, \dodoi{10.1103/RevModPhys.70.1}

\bibitem[{{Baronett} {et~al.}(2022){Baronett}, {Yang}, \& {Zhu}}]{2022BAAS...54e.381B}
{Baronett}, S.~A., {Yang}, C.-C., \& {Zhu}, Z. 2022, in Bulletin of the American Astronomical Society, Vol.~54, 102.381

\bibitem[{Bezanson {et~al.}(2017)Bezanson, Edelman, Karpinski, \& Shah}]{Julia}
Bezanson, J., Edelman, A., Karpinski, S., \& Shah, V.~B. 2017, SIAM {R}eview, 59, 65, \dodoi{10.1137/141000671}

\bibitem[{{Birnstiel} {et~al.}(2010){Birnstiel}, {Dullemond}, \& {Brauer}}]{2010AnA...513A..79B}
{Birnstiel}, T., {Dullemond}, C.~P., \& {Brauer}, F. 2010, \aap, 513, A79, \dodoi{10.1051/0004-6361/200913731}

\bibitem[{{Birnstiel} {et~al.}(2012){Birnstiel}, {Klahr}, \& {Ercolano}}]{2012A&A...539A.148B}
{Birnstiel}, T., {Klahr}, H., \& {Ercolano}, B. 2012, \aap, 539, A148, \dodoi{10.1051/0004-6361/201118136}

\bibitem[{{Brauer} {et~al.}(2008){Brauer}, {Dullemond}, \& {Henning}}]{Brauer08}
{Brauer}, F., {Dullemond}, C.~P., \& {Henning}, T. 2008, \aap, 480, 859, \dodoi{10.1051/0004-6361:20077759}

\bibitem[{{Carrera} \& {Simon}(2022)}]{2022BAAS...54e4502C}
{Carrera}, D., \& {Simon}, J. 2022, in Bulletin of the American Astronomical Society, Vol.~54, 405.02

\bibitem[{{Chen} \& {Lin}(2020)}]{Chen20}
{Chen}, K., \& {Lin}, M.-K. 2020, \apj, 891, 132, \dodoi{10.3847/1538-4357/ab76ca}

\bibitem[{{Dominik} \& {Dullemond}(2024)}]{2024A&A...682A.144D}
{Dominik}, C., \& {Dullemond}, C.~P. 2024, \aap, 682, A144, \dodoi{10.1051/0004-6361/202347716}

\bibitem[{{Dr{\.a}{\.z}kowska} \& {Dullemond}(2014)}]{D14}
{Dr{\.a}{\.z}kowska}, J., \& {Dullemond}, C.~P. 2014, \aap, 572, A78, \dodoi{10.1051/0004-6361/201424809}

\bibitem[{{Dra{\.z}kowska} {et~al.}(2019){Dra{\.z}kowska}, {Li}, {Birnstiel}, {Stammler}, \& {Li}}]{2019ApJ...885...91D}
{Dra{\.z}kowska}, J., {Li}, S., {Birnstiel}, T., {Stammler}, S.~M., \& {Li}, H. 2019, \apj, 885, 91, \dodoi{10.3847/1538-4357/ab46b7}

\bibitem[{{Dubrulle} {et~al.}(1995){Dubrulle}, {Morfill}, \& {Sterzik}}]{1995Icar..114..237D}
{Dubrulle}, B., {Morfill}, G., \& {Sterzik}, M. 1995, \icarus, 114, 237, \dodoi{10.1006/icar.1995.1058}

\bibitem[{{Estrada} {et~al.}(2016){Estrada}, {Cuzzi}, \& {Morgan}}]{2016ApJ...818..200E}
{Estrada}, P.~R., {Cuzzi}, J.~N., \& {Morgan}, D.~A. 2016, \apj, 818, 200, \dodoi{10.3847/0004-637X/818/2/200}

\bibitem[{{Estrada} {et~al.}(2022){Estrada}, {Cuzzi}, \& {Umurhan}}]{2022ApJ...936...42E}
{Estrada}, P.~R., {Cuzzi}, J.~N., \& {Umurhan}, O.~M. 2022, \apj, 936, 42, \dodoi{10.3847/1538-4357/ac7ffd}

\bibitem[{{Garate} {et~al.}(2020){Garate}, {Birnstiel}, {Drazkowska}, \& {Stammler}}]{2020A&A...635A.149G}
{Garate}, M., {Birnstiel}, T., {Drazkowska}, J., \& {Stammler}, S.~M. 2020, \aap, 635, A149, \dodoi{10.1051/0004-6361/201936067}

\bibitem[{Ho(2022)}]{MHDFlows}
Ho, K.~W. 2022, MHDFlows.jl, v.0.2.1b,  Zenodo, \dodoi{10.5281/zenodo.8242702}

\bibitem[{{Huang} \& {Bai}(2022)}]{AthenaDust}
{Huang}, P., \& {Bai}, X.-N. 2022, \apjs, 262, 11, \dodoi{10.3847/1538-4365/ac76cb}

\bibitem[{{Hunter}(2007)}]{2007Matplotlib}
{Hunter}, J.~D. 2007, Computing in Science and Engineering, 9, 90, \dodoi{10.1109/MCSE.2007.55}

\bibitem[{{Johansen} \& {Youdin}(2007)}]{Johansen07}
{Johansen}, A., \& {Youdin}, A. 2007, \apj, 662, 627, \dodoi{10.1086/516730}

\bibitem[{{Kowalik} {et~al.}(2013){Kowalik}, {Hanasz}, {W{\'o}lta{\'n}ski}, \& {Gawryszczak}}]{Kowalik13}
{Kowalik}, K., {Hanasz}, M., {W{\'o}lta{\'n}ski}, D., \& {Gawryszczak}, A. 2013, \mnras, 434, 1460, \dodoi{10.1093/mnras/stt1104}

\bibitem[{{Laibe} {et~al.}(2014){Laibe}, {Gonzalez}, {Maddison}, \& {Crespe}}]{2014MNRAS.437.3055L}
{Laibe}, G., {Gonzalez}, J.-F., {Maddison}, S.~T., \& {Crespe}, E. 2014, \mnras, 437, 3055, \dodoi{10.1093/mnras/stt1929}

\bibitem[{{Laune} {et~al.}(2020){Laune}, {Li}, {Li}, {Li}, {Walls}, {Birnstiel}, {Dr{\k{a}}{\.z}kowska}, \& {Stammler}}]{2020ApJ...889L...8L}
{Laune}, J., {Li}, H., {Li}, S., {et~al.} 2020, \apjl, 889, L8, \dodoi{10.3847/2041-8213/ab65c6}

\bibitem[{{Li} \& {Youdin}(2021)}]{LY21}
{Li}, R., \& {Youdin}, A.~N. 2021, \apj, 919, 107, \dodoi{10.3847/1538-4357/ac0e9f}

\bibitem[{{Li} {et~al.}(2020){Li}, {Li}, {Li}, {Birnstiel}, {Dr{\k{a}}{\.z}kowska}, \& {Stammler}}]{2020ApJ...892L..19L}
{Li}, Y.-P., {Li}, H., {Li}, S., {et~al.} 2020, \apjl, 892, L19, \dodoi{10.3847/2041-8213/ab7fb2}

\bibitem[{{Li} {et~al.}(2019){Li}, {Li}, {Li}, \& {Lin}}]{2019ApJ...886...62L}
{Li}, Y.-P., {Li}, H., {Li}, S., \& {Lin}, D. N.~C. 2019, \apj, 886, 62, \dodoi{10.3847/1538-4357/ab4bc8}

\bibitem[{{McNally} {et~al.}(2021){McNally}, {Lovascio}, \& {Paardekooper}}]{McNally21}
{McNally}, C.~P., {Lovascio}, F., \& {Paardekooper}, S.-J. 2021, \mnras, 502, 1469, \dodoi{10.1093/mnras/stab112}

\bibitem[{{National Academies of Sciences, Engineering, and Medicine}(2022)}]{nas2022origins}
{National Academies of Sciences, Engineering, and Medicine}. 2022, Origins, Worlds, and Life: A Decadal Strategy for Planetary Science and Astrobiology 2023-2032 (Washington, DC: The National Academies Press), \dodoi{10.17226/26522}

\bibitem[{{Rucska} \& {Wadsley}(2023)}]{Rucska23}
{Rucska}, J.~J., \& {Wadsley}, J.~W. 2023, \mnras, 526, 1757, \dodoi{10.1093/mnras/stad2855}

\bibitem[{{Schaffer} {et~al.}(2018){Schaffer}, {Yang}, \& {Johansen}}]{Schaffer18}
{Schaffer}, N., {Yang}, C.-C., \& {Johansen}, A. 2018, \aap, 618, A75, \dodoi{10.1051/0004-6361/201832783}

\bibitem[{{Sengupta} \& {Umurhan}(2023)}]{2023ApJ...942...74S}
{Sengupta}, D., \& {Umurhan}, O.~M. 2023, \apj, 942, 74, \dodoi{10.3847/1538-4357/ac9411}

\bibitem[{{Stone} {et~al.}(2020){Stone}, {Tomida}, {White}, \& {Felker}}]{Athena++}
{Stone}, J.~M., {Tomida}, K., {White}, C.~J., \& {Felker}, K.~G. 2020, \apjs, 249, 4, \dodoi{10.3847/1538-4365/ab929b}

\bibitem[{{Tominaga} \& {Tanaka}(2023)}]{Tominaga24}
{Tominaga}, R.~T., \& {Tanaka}, H. 2023, \apj, 958, 168, \dodoi{10.3847/1538-4357/ad002e}

\bibitem[{{Umurhan} {et~al.}(2020){Umurhan}, {Estrada}, \& {Cuzzi}}]{2020ApJ...895....4U}
{Umurhan}, O.~M., {Estrada}, P.~R., \& {Cuzzi}, J.~N. 2020, \apj, 895, 4, \dodoi{10.3847/1538-4357/ab899d}

\bibitem[{{Wurm} \& {Teiser}(2021)}]{2021NatRP...3..405W}
{Wurm}, G., \& {Teiser}, J. 2021, Nature Reviews Physics, 3, 405, \dodoi{10.1038/s42254-021-00312-7}

\bibitem[{{Yap} \& {Batygin}(2024)}]{2024Icar..41716085Y}
{Yap}, T.~E., \& {Batygin}, K. 2024, \icarus, 417, 116085, \dodoi{10.1016/j.icarus.2024.116085}

\bibitem[{{Youdin}(2010)}]{2010EAS....41..187Y}
{Youdin}, A.~N. 2010, in EAS Publications Series, Vol.~41, EAS Publications Series, ed. T.~{Montmerle}, D.~{Ehrenreich}, \& A.~M. {Lagrange}, 187--207, \dodoi{10.1051/eas/1041016}

\bibitem[{{Youdin} \& {Goodman}(2005)}]{SI}
{Youdin}, A.~N., \& {Goodman}, J. 2005, \apj, 620, 459, \dodoi{10.1086/426895}

\bibitem[{{Zhu} \& {Yang}(2021)}]{Zhu21}
{Zhu}, Z., \& {Yang}, C.-C. 2021, \mnras, 501, 467, \dodoi{10.1093/mnras/staa3628}

\end{thebibliography}



\end{document}